\documentclass[final,3p,times]{elsarticle}

\usepackage{amssymb}
\usepackage{amsmath}
\usepackage{amsthm} 
\usepackage{enumitem}
\usepackage{tabularx}
\newtheorem{theorem}{Theorem}

\newtheorem{observation}{Observation}

\usepackage{comment}
\usepackage{booktabs}
\usepackage{geometry}
\usepackage{rotating}
\usepackage{booktabs}
\usepackage{xcolor}

\journal{Operations Research Letters}

\begin{document}

\begin{frontmatter}



\title{Bundling and Price-Matching in Competitive Complementary Goods Markets}


\author[1]{Esmat Sangari\corref{cor1}} 
\ead{esangari@usfca.edu}
\author[2]{Rajni Kant Bansal} %
\ead{rajnikantb@iima.ac.in}
\affiliation[1]{organization={School of Management, University of San Francisco},
            addressline={2130 Fulton Street}, 
            city={San Francisco},
            postcode={94117}, 
            state={CA},
            country={United States}}
\affiliation[2]{organization={Area of Operations and Decision Sciences, Indian Institute of Management Ahmedabad},
addressline={Ahmedabad, GJ},
country={India}}
\cortext[cor1]{Corresponding author}

\begin{abstract}
We study mixed bundling and competitive price-matching guarantees (PMGs) in a duopoly selling complementary products to heterogeneous customers. One retailer offers mixed bundling while the rival sells only a bundle. We characterize unique pure-strategy Nash equilibria across subgames and compare them to a no-bundling benchmark. Mixed bundling strictly dominates whenever an equilibrium exists. Conditional on bundling, PMG adoption trades off strategic demand capture against margin losses on loyal customers and varies systematically with relative demand responsiveness to prices and complementarities.

\end{abstract}

\end{frontmatter}



\vspace{-1.5em}
\section{Introduction}\label{sec1}
\vspace{-0.5em}
\noindent
Bundling and price-matching guarantees (PMGs) are widely used pricing tools in markets where customers purchase complementary products. Retailers frequently sell curated bundles, such as laptops with accessories, home-office kits, or meal kits, to stimulate joint demand and increase overall profitability, as widely observed in industry practice. At the same time, major retailers employ PMGs to attract price-sensitive customers and maintain competitiveness. For example, Best Buy and Target publicly guarantee that they will match lower prices found at eligible competitors, both in-store and online.\\
[0.2cm]
While many firms offer bundles alongside stand-alone items, some retailers operate in a pure-bundling format, selling only bundled offerings. Examples include meal-kit services such as HelloFresh and Blue Apron, beauty-subscription services such as Birchbox and Ipsy, and all-inclusive travel providers that sell lodging, meals, and activities exclusively as packages.\\
[0.2cm]
These industry practices raise an important strategic question: How do bundling and price-matching interact when retailers compete in markets for complementary goods? Despite the widespread use of both strategies in practice, it remains unclear how PMGs influence the profitability of bundling, or how bundling affects the effectiveness of PMGs under heterogeneous customer behavior. Although the research streams on bundling and PMGs are well-developed, they have evolved largely in isolation and do not examine the interactions between these strategies. The bundling literature shows that bundling complementary goods can reshape competition and improve profitability (see, e.g., Giri et al., 2020; Chen et al., 2021; Chen et al., 2024) but does not incorporate PMGs. Conversely, the PMG literature analyzes price-matching strategies primarily in single-product settings (see, e.g., Zhuo, 2017; Jiang et al., 2017; Nalca \& Cai, 2023) and therefore does not examine bundling of complementary goods.\\
[0.2cm]
To the best of our knowledge, no prior study jointly analyzes bundling of complementary goods and competitive PMGs in a setting with heterogeneous customers and asymmetric retail formats. Our paper fills this gap by modeling a two-retailer duopoly, where one retailer (Retailer 1) chooses whether to offer mixed bundling or only stand-alone items, while the other retailer (Retailer 2) sells only a bundle. Conditional on this decision, each retailer chooses whether to adopt a PMG and sets prices. Customers fall into three groups, namely, loyal price-unaware, loyal price-aware, and non-loyal price-aware (strategic) customers, capturing realistic purchasing behavior observed in practice. \\
[0.2cm]
Our equilibrium analysis shows that under mixed bundling with a PMG, Retailer 1’s bundle price (and, through bundle-item interactions, its individual-item prices) depend on Retailer 2’s bundle-demand parameters, because the effective bundle price faced by price-aware customers is tied to the rival’s price. In contrast, Retailer 2’s bundle price is determined solely by its own demand parameters. When Retailer 1 does not adopt a PMG under mixed bundling, or when its equilibrium bundle price is lower than Retailer 2’s bundle price, its prices depend only on its own demand parameters. These results clarify when the interaction between mixed bundling and PMGs transmits competitive conditions across retailers through the effective bundle price.\\
[0.2cm]
Our numerical experiments map the profitability of bundling and PMG adoption across demand environments. Mixed bundling strictly dominates stand-alone selling whenever an equilibrium exists, with the largest gains arising when bundle complementarity is sufficiently high and stand-alone interaction effects are relatively weak. Conditional on bundling, PMGs are unattractive when strategic customers are highly price-sensitive and loyal customers are relatively price-insensitive, but become optimal when both groups exhibit medium-to-high price-sensitivity. Equilibrium is less likely under strong asymmetries in price-sensitivity across customer groups and more robust when bundle complementarity is high and stand-alone interaction effects are limited.\\
[0.2cm]
Overall, our findings provide the first integrated characterization of how bundling and PMGs jointly shape competition for complementary goods and offer actionable guidance to retailers facing bundle-only competitors or strategically considering PMG adoption.

\vspace{-1em}
\section{Literature Review}\label{sec2}
\vspace{-0.5em}
\noindent
Our study lies at the intersection of two well-established research streams: bundling of complementary products and competitive PMGs. Although both streams have been extensively studied in \mbox{isolation}, their intersection remains largely underexplored. We contribute a unified model that jointly analyzes mixed bundling and price-matching decisions in a competitive setting with heterogeneous consumers.\\
[0.2cm]
In the bundling literature, prior work has explored pure and mixed bundling strategies across supply chain \linebreak{structures}, demand types, and product relationships. For complementary goods, bundling is often shown to \mbox{enhance} total profits under various channel power arrangements and product complementarities (Giri et al., 2020; Chen et al., 2021). Competitive bundling has been studied under multi-product settings (Liao \& Tauman, 2002; De Cornière \& Taylor, 2024), and in duopolies with differentiated complements (Chung et al., 2013; Vamosiu, 2018). The literature has also extended bundling analysis to settings involving customers’ brand-matching preferences (Sinitsyn, 2012), green product strategies (Shan et al., 2020), and omni-channel supply chains (Chen et al., 2024). Some recent work further explores bundling considering advertising dynamics under varying channel power (Jena \& Ghadge, 2022) and supplier risk aversion and stochastic demand (Hemmati et al., 2024). However, none of these studies incorporate PMGs into bundling decisions.\\
[0.2cm]
In parallel, the literature on PMGs has examined their strategic and welfare implications across various \mbox{market} \mbox{settings}. While early work raised concerns that PMGs may facilitate tacit collusion (Corts, 1996; Zhuo, 2017), later studies highlight countervailing effects depending on consumer behavior and operational context. For \linebreak{instance}, PMGs may heighten competition by encouraging consumer search (Chen et al., 2001; Jiang et al., 2017), or enable price discrimination under availability contingencies (Nalca et al., 2010; Nalca et al., 2019). The profitability of PMGs further depends on market structure, channel power, demand asymmetries, and hassle costs (Nalca, 2017; Nalca \& Cai, 2023). Empirical evidence offers mixed findings, with some studies supporting pro-competitive interpretations (Wu et al., 2018), and others documenting price increases even in non-adopting rivals (Zhuo, 2017). More recent studies examine how PMGs interact with logistics conditions (Wei \& Chang, 2023) and dual-channel dynamics shaped by product differentiation and price-sensitivity (Wei et al., 2023).\\
[0.2cm]
Despite this rich body of work, we are not aware of any study that explicitly models how bundling and price-matching strategies interact for complementary products in competitive retail markets. Our paper fills this gap by examining how a retailer’s decision to offer a mixed bundle or separate products is shaped by both its own and its rival’s PMG decisions. By linking these two strategic levers in a common game-theoretic framework, we provide new insights into competitive dynamics, demand segmentation, and firm profitability in retail environments.

\vspace{-1em}
\section{Model Setup}\label{sec3}
\vspace{-0.5em}
\noindent
In this section, we define the market, firms’ decisions, consumer segments, demand structure, and profit functions.\\
[0.2cm]
{\bf\small Duopoly Competition and Firms' Decisions:} We study a market with two retailers selling two complementary \mbox{products}. Retailer 1 ($r_1$) chooses whether to offer mixed bundling ($B = 1$), setting prices $p_{r1}^{i1}$, $p_{r1}^{i2}$, and $p_{r1}^b$ for individual items 1 and 2 and the bundle, respectively, or not bundle ($B = 0$), where it sets only individual prices $p_{r1}^{i1}$ and $p_{r1}^{i2}$. Retailer 2 ($r_2$) sells bundles only at price $p_{r2}^b$. Both firms decide whether to offer a competitive PMG and set prices. Product costs are $c_1$ and $c_2$ for items 1 and 2, respectively. We impose $p_{r1}^{i1} + p_{r1}^{i2} \geq p_{r1}^b$, ensuring that the bundle price does not exceed the combined price of its components. Figure 1 illustrates the overall decision sequence in this duopoly game.\\
[0.2cm]
\begin{figure}[b]
\vspace{-1em}
\centering
\includegraphics[width=0.79\linewidth]{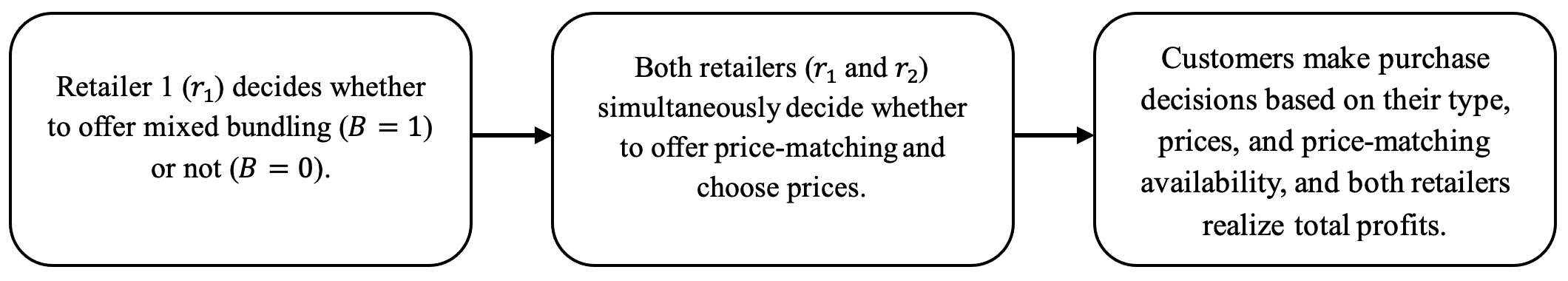}
\vspace{-0.8em}
\caption{Sequence of Decisions and Interactions in the Duopoly Game}
\label{game}
\end{figure}
{\bf\small Customer Segmentation:} Customers are partitioned into three behaviorally distinct segments: 
\begin{itemize}[leftmargin=*]
\vspace{-0.5em}
    \item \textit{Loyal, price-unaware customers}, who consistently buy from their preferred retailer without comparing prices. Within this segment, some purchase only one individual item from $r_1$, while others purchase both products (either individually or as a bundle if offered). Customers loyal to $r_2$ can only purchase the bundle, since this retailer does not sell individual items.
\vspace{-0.5em}
    \item \textit{Loyal, price-aware customers}, who remain committed to their retailer but take advantage of PMGs when available. We assume these customers always purchase both products (as a bundle if offered, or as \mbox{individual} items otherwise), reflecting that price awareness is only relevant when considering the joint \mbox{purchase}.
\vspace{-0.5em}
    \item \textit{Non-loyal, price-aware (i.e., strategic) customers}, who actively compare the price of \mbox{obtaining} both products across retailers. They evaluate the bundle price when offered (or the sum of individual prices when bundling is not available at $r_1$) and purchase from the retailer offering the lowest effective price. 
\end{itemize}
\vspace{0em}
Accordingly, single-item purchases arise only among price-unaware customers, while all price-aware customers evaluate the joint purchase, making PMGs relevant only at the bundle level because the bundle is the only cross-retailer comparable offer.\\
[0.2cm]
{\bf\small Demand Functions:} We assume linear demands with own-price sensitivity $b_l > 0$, complementarity \mbox{parameters} $\theta_l > 0$ (across individual items) and $\lambda_l > 0$ (between the bundle and the sum of parts), and segment-specific demand bases $a_l^v$ and $a_q^v$ for $v \in \{i1,i2,ib,jb\}$, where $l$ and $q$ denote loyal, price-unaware and loyal, price-aware customers, respectively, $i1$, $i2$, and $ib$ denote the individual items 1 and 2 and the bundle in $r_1$, respectively, and $jb$ denotes the bundle offered by $r_2$. In addition, $a_s$ denotes the demand base of strategic customers. We assume $b_l \ge \lambda_l$, so that demand is more sensitive to changes in a product’s own price than to the price difference between the bundle and separate purchases. In particular: 
\begin{itemize}[leftmargin=*]
\item{\it Demand Function of Loyal, Price-Unaware Customers}: When $B = 1$ (i.e., mixed bundling at $r_1$), $r_1$'s individual-item and bundle demands in this segment are, respectively, as follows:
\vspace{-0.5em}
\begin{align}
& \!\!\!\!\!\!\!d_l^{i1}(p_{r_1}^{i1},p_{r_1}^{i2},p_{r_1}^b) \!=\! a_l^{i1} \!\!-\! b_l p_{r_1}^{i1} \!-\! b_l \theta_l p_{r_1}^{i2} \!+\! \lambda_l(p_{r_1}^{b} \!-\! p_{r_1}^{i1} \!-\! p_{r_1}^{i2}), \ \ d_l^{i2}(p_{r_1}^{i1},p_{r_1}^{i2},p_{r_1}^b) \!=\! a_l^{i2} \!\!-\! b_l \theta_lp_{r_1}^{i1} \!-\! b_l p_{r_1}^{i2} \!+\! \lambda_l(p_{r_1}^b \!-\! p_{r_1}^{i1} \!-\!p_{r_1}^{i2}) \\
& \!\!\!\!\!\!\!d_l^{ib}(p_{r_1}^{i1},p_{r_1}^{i2},p_{r_1}^b) = a_l^{ib} - b_l p_{r_1}^b + \lambda_l(p_{r_1}^{i1} + p_{r_1}^{i2} - p_{r_1}^b)
\end{align}
And $r_2$'s bundle demand in this segment is as follows:
\vspace{-0.5em}
\begin{align}
d_l^{jb}(p_{r_2}^b) = a_l^{jb} - b_l p_{r_2}^b
\end{align}
When $B = 0$ (i.e., no bundling by $r_1$), the loyal, price-unaware demands at $r_1$ simplify to the following because $r_1$ does not post a bundle price, and $r_2$'s bundle demand in this segment remains unchanged:
\vspace{-0.5em}
\begin{align}
& \!\!\!\!\!\!\!d_l^{i1}(p_{r_1}^{i1},p_{r_1}^{i2}) \!=\! a_l^{i1} \!\!-\! b_l p_{r_1}^{i1} \!-\! b_l \theta_l p_{r_1}^{i2},  \ d_l^{i2}(p_{r_1}^{i1},p_{r_1}^{i2}) \!=\! a_l^{i2} \!\!-\! b_l \theta_lp_{r_1}^{i1} \!-\! b_l p_{r_1}^{i2} \\
& \!\!\!\!\!\!\!d_l^{ib}(p_{r_1}^{i1},p_{r_1}^{i2}) = a_l^{ib} - b_l(p_{r_1}^{i1} + p_{r_1}^{i2})\label{eq7}
\end{align}
In equation (\ref{eq7}), $d_l^{ib}$ represents the demand of those customers who need to purchase both items regardless of the bundling decision at $r_1$, reflecting the ``would-be'' bundle price at $r_1$ evaluated at sum of the two items' prices.
\item {\it Demand Function of Loyal, Price-Aware Customers}: Let the effective bundle price under PMGs for retailer $r$ be $\tilde{p}_r^b$. The PMG mapping for the effective prices is case-based: if PMGs are offered at $r_1$, and $p_{r1}^b \geq p_{r2}^b$, then $\tilde{p}_r^b = p_{r2}^b$, and vice versa for $r_2$; if a retailer does not offer a PMG, its effective bundle price coincides with its posted bundle price, i.e., $\tilde{p}_r^b = p_r^b$. Therefore, with mixed bundling at $r_1$, the bundle demand in this segment at $r_1$ and $r_2$ are, respectively, as follows:
\vspace{-0.5em}
\begin{align}
& \!\!\!\!\!\!\!\!d_q^{ib}(p_{r1}^{i1},p_{r1}^{i2},\Tilde{p}_{r_1}^b) \!=\! a_q^{ib} - b_l \Tilde{p}_{r_1}^b  + \lambda_l(p_{r1}^{i1} + p_{r1}^{i2}- \Tilde{p}_{r_1}^b)\\
& \!\!\!\!\!\!\!\!d_q^{jb}(\Tilde{p}_{r_2}^b) = a_q^{jb} - b_l \Tilde{p}_{r_2}^b
\end{align}
Note that, with no mixed bundling at $r_1$, no PMGs apply for this segment, and they evaluate the composite price. Accordingly, the bundle demand at $r_1$ and $r_2$ in this case are, respectively, as follows:
\vspace{-0.5em}
\begin{align}
& \!\!\!\!\!\!\!\!d_q^{ib}(p_{r1}^{i1},p_{r1}^{i2}) \!=\! a_q^{ib} - b_l(p_{r1}^{i1} + p_{r1}^{i2})\\
& \!\!\!\!\!\!\!\!d_q^{jb}(p_{r_2}^b) = a_q^{jb} - b_l p_{r_2}^b
\end{align}
\item {\it Demand Function of Non-Loyal, Price-Aware (Strategic) Customers}: Strategic customers buy from the lowest available bundle price. Accordingly, the demand in this section is: 
\vspace{-0.5em}
\begin{align}
\!\!\!\!\!\!\!\!d_s(\hat{p}^b) \!=\! a_s - b_s\hat{p}^b 
\end{align}
With mixed bundling at $r_1$, $\hat{p}^b = \min\{p_{r1}^b, p_{r2}^b\}$, and with no bundling at $r_1$, $\hat{p}^b = \min\{p_{r1}^{i1} + p_{r1}^{i2}, p_{r2}^b\}$.
\end{itemize}
{\bf\small Profit Functions:} Given the above demands, retailers maximize their profit by deciding on PMG adoption, when applicable, and by setting prices. When $B = 1$, the profit functions for $r_1$ and $r_2$ are, respectively, as follows:
\vspace{0em}
\begin{align}
& \!\!\!\pi_{r1}(p_{r1}^{i1},p_{r1}^{i2},p_{r1}^b,\Tilde{p}_{r_1}^b, \hat{p}^b) \!=\! (p_{r1}^{i1} \!-\! c_1)d_l^{i1} \!+\! (p_{r1}^{i2} \!-\! c_2)d_l^{i2} \!+\!  (p_{r1}^b \!-\! c_1 \!-\! c_2)d_l^{ib} \!+\! (\tilde{p}_{r_1}^b \!-\! c_1 \!-\! c_2)d_q^{ib} \!+\!  \alpha (\hat{p}^b\!-\!c_1\!-\!c_2)d_s \\
& \!\!\!\pi_{r2}(p_{r2}^b,\Tilde{p}_{r_2}^b, \hat{p}^b) = (p_{r2}^{b} \!-\! c_1 \!-\!c_2)d_l^{jb} \!+\! (\tilde{p}_{r_2}^b \!-\! c_1 \!-\! c_2)d_q^{jb} \!+\! (1\!-\!\alpha) (\hat{p}^b - c_1 - c_2)d_s
\end{align}
where $\alpha \in [0,1]$ denotes the proportion of strategic customers purchasing from $r_1$, based on the retailers’ pricing and PMG decisions.\\
[0.2cm]
When $B = 0$, the profit function for $r_2$ remains unchanged and the profit function for $r_1$ becomes as follows:
\begin{align}
& \pi_{r1}(p_{r1}^{i1},p_{r1}^{i2}, \hat{p}^b) = (p_{r1}^{i1} \!- c_1)d_l^{i1} + (p_{r1}^{i2} - c_2)d_l^{i2} +  (p_{r1}^{i1} + p_{r1}^{i2} \!-\! c_1 \!-\! c_2)(d_l^{ib} + d_q^{ib}) + \alpha (\hat{p}^b\!-\!c_1\!-\!c_2)d_s
\end{align}
{\bf\small Strategic Scenarios:} After $r_1$ makes its bundling decision ($B = 1$ or $B = 0$), both retailers simultaneously decide whether to offer PMGs (when applicable) and set their prices. This yields four possible strategic configurations under $B = 1$: $(\text{CM}, \text{CM})$, $(\overline{\text{CM}}, \text{CM})$, $(\text{CM}, \overline{\text{CM}})$, and $(\overline{\text{CM}}, \overline{\text{CM}})$, where $\text{CM}$ denotes the adoption of PMGs, and $\overline{\text{CM}}$ denotes no PMGs.

\vspace{-1em}
\section{Equilibrium Analysis}\label{sec4}
\vspace{-0.5em}
\noindent
In this section, we characterize the closed-form equilibrium solutions for all four bundling subgames and the benchmark case without bundling. We derive the equilibria by analyzing the KKT conditions of each firm's profit-maximization problem and solving them simultaneously.
\begin{theorem}\label{asymmetric_eqbm_1_clean}
Under Sufficient Condition Set A (see Appendix A), the profit functions 
$\pi_{r1}(p_{r1}^{i1},p_{r1}^{i2},p_{r1}^b,p_{r2}^{b})$ and 
$\pi_{r2}(p_{r1}^b,p_{r2}^{b})$ are jointly concave in subgames 
$(\text{CM},\text{CM})$ and $(\text{CM},\overline{\text{CM}})$ when $r_1$ adopts mixed bundling (i.e., $B=1$). 
A unique pure-strategy Nash equilibrium exists in these subgames as follows:
\vspace{-0.5em}
 \begin{subequations}
            \begin{align}
                & \!\!\! p_{r1}^{i1} \!=\! \frac{1}{4}\frac{a_l^{i1}\!-\!a_l^{i2}}{b_l(1\!-\!\theta_l)} \!-\! \frac{a_l^{ib}}{4\lambda_l} \!+\!  \frac{2p_{r1}^b \!-\! c_1 \!-\! c_2}{4\lambda_l}(b_l\!+\!\lambda_l) \!+\!  \frac{c_1}{2} \label{case_opt_sol.a}\\ 
                & \!\!\! p_{r1}^{i2} \!=\! -\frac{1}{4}\frac{a_l^{i1} \!-\!a_l^{i2}}{b_l(1\!-\!\theta_l)} \!-\! \frac{a_l^{ib}}{4\lambda_l} \!+\!  \frac{2p_{r1}^b \!-\! c_1 \!-\! c_2}{4\lambda_l}(b_l\!+\!\lambda_l) \!+\!  \frac{c_2}{2} \label{case_opt_sol.b}\\
                & \!\!\!p_{r1}^b \!=\! \frac{1}{2}\left(\frac{(a_l^{i1}\!+\!a_l^{i2})\lambda_l\!+\!(b_l(1\!+\!\theta_l)\!+\!2\lambda_l)a_l^{ib}}{(b_l+\lambda_l)b_l(1+\theta_l)+2b_l\lambda_l} \!+\! \left(\frac{a_l^{jb}\!+\! a_q^{jb} \!+\! (1\!-\!\alpha)a_s}{2b_l+(1-\alpha)b_s} \!-\! (c_1\!+\!c_2)\!\right)\frac{(\lambda_l)^2}{(b_l\!+\!\lambda_l)b_l(1\!+\!\theta_l)\!+\!2b_l\lambda_l}\!+\! c_1\!+\!c_2\right) \label{case_opt_sol.c}\\
                & \!\!\!p_{r2}^{b} = \frac{1}{2}\left(\frac{a_l^{jb}\!+\! a_q^{jb} \!+\! (1\!-\!\alpha)a_s}{2b_l\!+\!(1\!-\!\alpha)b_s} + (c_1 \!+\! c_2)\right)  \label{case_opt_sol.d}
            \end{align}
           \end{subequations}
\end{theorem}
\vspace{-0.5em}
\noindent
The equilibrium in Theorem \ref{asymmetric_eqbm_1_clean} exhibits an important asymmetry. The optimal bundle price of $r_1$ (i.e., $p_{r1}^{b}$) increases with $r_2$'s bundle-related base demands ($a_l^{jb}, a_q^{jb}$). Moreover, the sensitivity of this response increases with $\lambda_l$, since the competition term in $p_{r1}^{b}$ is scaled by $\lambda_l^2$. The price dependence for $r_1$ occurs because, in this case, $r_1$ offers a PMG and its effective bundle price for price-aware customers is matched to $r_2$'s lower bundle price. In contrast, $r_2$'s bundle price (i.e., $p_{r2}^{b}$) depends only on its own demand parameters and does not vary with $r_1$'s individual or bundle demands. Note that under mixed bundling with a PMG at $r_1$, its individual-item prices depend indirectly on the rival’s demand parameters through its equilibrium bundle price.  
\begin{theorem}\label{asymmetric_eqbm_2_clean}
        Under Sufficient Condition Set B (See Appendix B), the profit functions $\pi_{r1}(p_{r1}^{i1},p_{r1}^{i2},p_{r1}^b,p_{r2}^{b})$ and $ \pi_{r2}(p_{r1}^b,p_{r2}^{b})$ are jointly concave in subgames $(\overline{\text{CM}}, \text{CM})$ and $(\overline{\text{CM}}, \overline{\text{CM}})$ when $r_1$ adopts mixed bundling (i.e., $B = 1$). A unique pure-strategy Nash equilibrium exists in these subgames as follows:
        \vspace{-0.5em}
           \begin{subequations}
            \begin{align}
                & \!\!p_{r1}^{i1} \!=\! \frac{1}{4}\frac{a_l^{i1} \!-\!a_l^{i2}}{b_l(1 \!-\!\theta_l)} \!+\! \frac{5c_1}{12} \!-\!  \frac{c_2}{12} \!-\! \frac{a_l^{ib}\!+\! a_q^{ib}}{6\lambda_l}  \!+\!  \frac{2p_{r1}^b \!-\! c_1 \!-\! c_2}{3\lambda_l}(b_l\!+\!\lambda_l) \\
                & \!\!p_{r1}^{i2} \!=\! -\frac{1}{4}\frac{a_l^{i1}\!-\!a_l^{i2}}{b_l(1\!-\!\theta_l)} \!-\! \frac{c_1}{12} \!+\!  \frac{5c_2}{12} \!-\! \frac{a_l^{ib}\!+\! a_q^{ib}}{6\lambda_l}  \!+\!  \frac{2p_{r1}^b \!-\! c_1 \!-\! c_2}{3\lambda_l}(b_l\!+\!\lambda_l) \\
                & \!\!p_{r1}^b = \frac{1}{2}\left(\!\frac{3(a_l^{i1}+a_l^{i2})\lambda_l+ 2(b_l(1+\theta_l)+2\lambda_l)(a_l^{ib}+ a_q^{ib})}{4b_l(b_l(1+\theta_l)+2\lambda_l) + 4b_l(1+\theta_l)\lambda_l -   (\lambda_l)^2} \!+ \!(c_1 \!+\! c_2)\! \left( \!1 \!+\! \frac{b_l(1+\theta_l)\lambda_l - (\lambda_l)^2}{4b_l(b_l(1\!+\!\theta_l)\!+\!\!2\lambda_l) \!\!+\! 4b_l(1\!+\!\theta_l)\lambda_l \!-\!   (\lambda_l)^2}\!\right)\right) \\
                & \!\! p_{r2}^{b} = \frac{1}{2}\left(\frac{a_l^{jb}+ a_q^{jb} + a_s}{2b_l+b_s} + (c_1 + c_2)\right)
            \end{align}
           \end{subequations}
\end{theorem}
\vspace{-0.5em}
\noindent
In Theorem \ref{asymmetric_eqbm_2_clean}, $r_2$'s bundle price $p_{r2}^{b}$ depends only on its own demand parameters and is independent of $r_1$'s individual or bundle demand parameters. Additionally, in contrast to Theorem~\ref{asymmetric_eqbm_1_clean}, when $r_1$ does not offer a PMG, its optimal bundle price $p_{r1}^{b}$, and hence its individual-item prices, depend only on its own demand parameters.  
\begin{theorem}\label{asymmetric_eqbm_1_flip_clean}
        Under Sufficient Condition Set C (See Appendix C), the profit functions $\pi_{r1}(p_{r1}^{i1},p_{r1}^{i2},p_{r1}^b,p_{r2}^{b})$ and $ \pi_{r2}(p_{r1}^b,p_{r2}^{b})$ are jointly concave in subgames $({\text{CM}}, {\text{CM}})$ and $(\overline{\text{CM}},{\text{CM}})$ when $r_1$ adopts mixed bundling (i.e., $B = 1$). A unique pure-strategy Nash equilibrium exists in these subgames as follows:
        \vspace{-0.5em}
           \begin{subequations}
            \begin{align}
                & \!\!p_{r1}^{i1} \!=\! \frac{1}{4}\frac{a_l^{i1} \!-\!a_l^{i2}}{b_l(1 \!-\!\theta_l)} \!+\! \frac{5c_1}{12} \!-\!  \frac{c_2}{12} \!-\! \frac{a_l^{ib}\!+\! a_q^{ib}+\alpha a_s}{6\lambda_l}  \!+\!  \frac{2p_{r1}^b \!-\! c_1 \!-\! c_2}{3\lambda_l}(b_l\!+\!\lambda_l+\frac{\alpha}{2}b_S) \\
                & \!\!p_{r1}^{i2} \!=\! -\frac{1}{4}\frac{a_l^{i1}\!-\!a_l^{i2}}{b_l(1\!-\!\theta_l)} \!-\! \frac{c_1}{12} \!+\!  \frac{5c_2}{12} \!-\! \frac{a_l^{ib}\!+\! a_q^{ib}+\alpha a_s}{6\lambda_l}  \!+\!  \frac{2p_{r1}^b \!-\! c_1 \!-\! c_2}{3\lambda_l}(b_l\!+\!\lambda_l+\frac{\alpha}{2}b_s) \\
                & \!\!p_{r1}^b \!= \!\frac{1}{2}\!\left(\!\frac{3(a_l^{i1}\!\!+\!a_l^{i2})\lambda_l+ 2(b_l(1\!+\!\theta_l)\!+\!2\lambda_l)(a_l^{ib}\!+\! a_q^{ib})\!+\!\alpha a_s}{2(2b_l\!+\!\alpha b_s)(b_l(1\!\!+\!\theta_l)\!\!+\!2\lambda_l) \!\!+\! 4b_l(1\!\!+\!\theta_l)\lambda_l \!-\!   (\lambda_l)^2} \! + \!(c_1 \!+\! c_2)\! \left( \!1 \!+\! \frac{b_l(1\!+\!\theta_l)\lambda_l - (\lambda_l)^2}{2(2b_l\!+\!\alpha b_s)(b_l(1\!\!+\!\theta_l)\!+\!\!2\lambda_l) \!\!+\! 4b_l(1\!\!+\!\theta_l)\lambda_l \!-\!   (\lambda_l)^2}\!\right)\right) \\
                & \!\! p_{r2}^{b} = \frac{1}{2}\left(\frac{a_l^{jb}}{b_l} + (c_1 + c_2)\right)
            \end{align}
           \end{subequations}
\end{theorem}
\vspace{-0.5em}
\noindent
In Theorem~\ref{asymmetric_eqbm_1_flip_clean}, the equilibrium satisfies $p_{r1}^{b} <  p_{r2}^{b}$. A key property here is that, if $\alpha=0$, $r_1$’s equilibrium prices coincide with those in Theorem~\ref{asymmetric_eqbm_2_clean}. This implies that, in the absence of strategic customers, $r_1$’s PMG adoption does not affect its optimal prices, since the remaining customer segments do not switch to $r_2$. More generally, $r_1$'s optimal bundle price $p_{r1}^{b}$ depends only on its own demand parameters, regardless of whether it offers a PMG. Since $r_2$ offers a PMG in this case, its bundle price $p_{r2}^{b}$ is determined solely by demand from loyal, price-unaware customers, represented by $a_l^{jb}$, as loyal, price-aware customers purchase at the matched lower price offered by $r_1$.
\begin{theorem}\label{asymmetric_eqbm_2_flip_clean}
        Under Sufficient Condition Set D (See Appendix D), the profit functions $\pi_{r1}(p_{r1}^{i1},p_{r1}^{i2},p_{r1}^b,p_{r2}^{b})$ and $ \pi_{r2}(p_{r1}^b,p_{r2}^{b})$ are jointly concave in subgames $({\text{CM}}, \overline{\text{CM}})$ and $(\overline{\text{CM}}, \overline{\text{CM}})$ when $r_1$ adopts mixed bundling (i.e., $B = 1$). A unique pure-strategy Nash equilibrium exists in these subgames as follows:
        \vspace{-0.5em}
           \begin{subequations}
            \begin{align}
                & \!\!p_{r1}^{i1} \!=\! \frac{1}{4}\frac{a_l^{i1} \!-\!a_l^{i2}}{b_l(1 \!-\!\theta_l)} \!+\! \frac{5c_1}{12} \!-\!  \frac{c_2}{12} \!-\! \frac{a_l^{ib}\!+\! a_q^{ib}+ a_s}{6\lambda_l}  \!+\!  \frac{2p_{r1}^b \!-\! c_1 \!-\! c_2}{3\lambda_l}(b_l\!+\!\lambda_l+\frac{1}{2}b_S) \\
                & \!\!p_{r1}^{i2} \!=\! -\frac{1}{4}\frac{a_l^{i1}\!-\!a_l^{i2}}{b_l(1\!-\!\theta_l)} \!-\! \frac{c_1}{12} \!+\!  \frac{5c_2}{12} \!-\! \frac{a_l^{ib}\!+\! a_q^{ib}+ a_s}{6\lambda_l}  \!+\!  \frac{2p_{r1}^b \!-\! c_1 \!-\! c_2}{3\lambda_l}(b_l\!+\!\lambda_l+\frac{1}{2}b_s) \\
                & \!\!p_{r1}^b \!=\! \frac{1}{2}\!\left(\!\frac{3(a_l^{i1}\!+\!a_l^{i2})\lambda_l\!+ 2(b_l(1\!\!+\!\theta_l)\!+\!2\lambda_l)(a_l^{ib}\!+\! a_q^{ib})\!+\! a_s}{2(2b_l\!+\! b_s)(b_l(1\!\!+\!\theta_l)\!+\!2\lambda_l) \!+\! 4b_l(1\!+\!\theta_l)\lambda_l \!-\!   (\lambda_l)^2} \!+ \!(c_1 \!+\! c_2)\! \left( \!1 \!+\! \frac{b_l(1\!\!+\!\theta_l)\lambda_l \!-\! (\lambda_l)^2}{2(2b_l\!+\! b_s)(b_l(1\!\!+\!\theta_l)\!+\!\!2\lambda_l) \!\!+\! 4b_l(1\!\!+\!\theta_l)\lambda_l \!-\!   (\lambda_l)^2}\!\right)\right) \\
                & \!\! p_{r2}^{b} = \frac{1}{2}\left(\frac{a_l^{jb}+a_q^{jb}}{2b_l} + (c_1 + c_2)\right)
            \end{align}
           \end{subequations}
\end{theorem}
\vspace{-0.5em}
\noindent
In Theorem~\ref{asymmetric_eqbm_2_flip_clean}, $r_1$’s equilibrium prices coincide with those in Theorem~\ref{asymmetric_eqbm_1_flip_clean} when $\alpha=1$, implying that $r_1$ captures strategic customers when $p_{r1}^{b} < p_{r2}^{b}$ and $r_2$ does not offer a PMG. Moreover, as in Theorem~\ref{asymmetric_eqbm_1_flip_clean}, $r_1$’s bundle price $p_{r1}^{b}$ depends only on its own demand parameters. Since $r_2$ does not offer a PMG in this case, its bundle price $p_{r2}^{b}$ is determined by both loyal, price-aware and loyal, price-unaware demand, represented by $a_q^{jb}$ and $a_l^{jb}$. 
\begin{theorem}\label{asymmetric_eqbm_no_bundle_clean}
When $r_1$ sells only individual items (i.e., $B = 0$), under each of Sufficient Condition Sets E and F (see Appendix E), the profit functions $\pi_{r1}(p_{r1}^{i1},p_{r1}^{i2},p_{r1}^b,p_{r2}^{b})$ and 
$\pi_{r2}(p_{r1}^b,p_{r2}^{b})$ are jointly concave.
Under each condition set, a unique pure-strategy Nash equilibrium exists as follows:\\
[0.2cm]
{(a) Under Sufficient Condition Set E:}
\vspace{-0.5em}
\begin{subequations}
\begin{align}
& p_{r1}^{i1} \!=\! \frac{a_l^{i1} \!-\! a_l^{i2}}{4b_l(1\!-\!\theta_l)} \!+\! \frac{c_1}{2} \!+\! \frac{a_l^{i1} \!+\! a_l^{i2} \!+\! 2(a_l^{ib} \!+\! a_q^{ib})}{4b_l(5+\theta_l)}\\
& p_{r1}^{i2} \!=\! - \frac{a_l^{i1} \!-\! a_l^{i2}}{4b_l(1\!-\!\theta_l)} \!+\! \frac{c_2}{2} \!+\! \frac{a_l^{i1} \!+\! a_l^{i2} \!+\! 2(a_l^{ib} \!+\! a_q^{ib})}{4b_l(5+\theta_l)}\\
& p_{r2}^{b} = \frac{a_l^{jb}+a_q^{jb} \!+\! a_s}{2(2b_l+b_s)}+\frac{1}{2}(c_1+c_2) 
\end{align}
\end{subequations}
{(b) Under Sufficient Condition Set F:}
\vspace{-0.5em}
\begin{subequations}
\begin{align}
& p_{r1}^{i1} \!=\! \frac{a_l^{i1} \!-\! a_l^{i2}}{4b_l(1\!-\!\theta_l)} \!+\! \frac{c_1}{2} \!+\! \frac{a_l^{i1} \!+\! a_l^{i2} \!+ a_s +\! 2(a_l^{ib} \!+\! a_q^{ib})}{4b_l(5+\theta_l)+4b_s}\\
& p_{r1}^{i2} \!=\! - \frac{a_l^{i1} \!-\! a_l^{i2}}{4b_l(1\!-\!\theta_l)} \!+\! \frac{c_2}{2} \!+\! \frac{a_l^{i1} \!+\! a_l^{i2} \!+ a_s + \! 2(a_l^{ib} \!+\! a_q^{ib})}{4b_l(5+\theta_l)+4b_s}\\
& p_{r2}^{b} = \frac{a_l^{jb}+a_q^{jb}}{4b_l}+\frac{1}{2}(c_1+c_2) 
\end{align}
\end{subequations}
\end{theorem}
\vspace{-0.5em}
\noindent
In Theorem~\ref{asymmetric_eqbm_no_bundle_clean}, when $r_1$ does not offer a bundle, the bundle-price channel linking $r_1$’s prices to the rival's price is eliminated. Under Sufficient Condition Set E, $r_1$’s individual-item prices depend only on its own demand bases and cost parameters, in contrast to the mixed-bundling case with a PMG (Theorem~\ref{asymmetric_eqbm_1_clean}), where cross-retailer dependence arises through the equilibrium bundle price. Under Sufficient Condition Set F, strategic customers are captured by $r_1$ since $p_{r1}^{i1}+p_{r1}^{i2}<p_{r2}^{b}$, while $r_2$’s bundle price is driven solely by its loyal demand. In both cases, the absence of bundling implies that $r_1$’s prices depend exclusively on its own demand parameters.

\vspace{-1em}
\section{Numerical Experiments}\label{sec5}
\begin{figure*}[t]
    \centering 
    \includegraphics[width=0.81\linewidth]{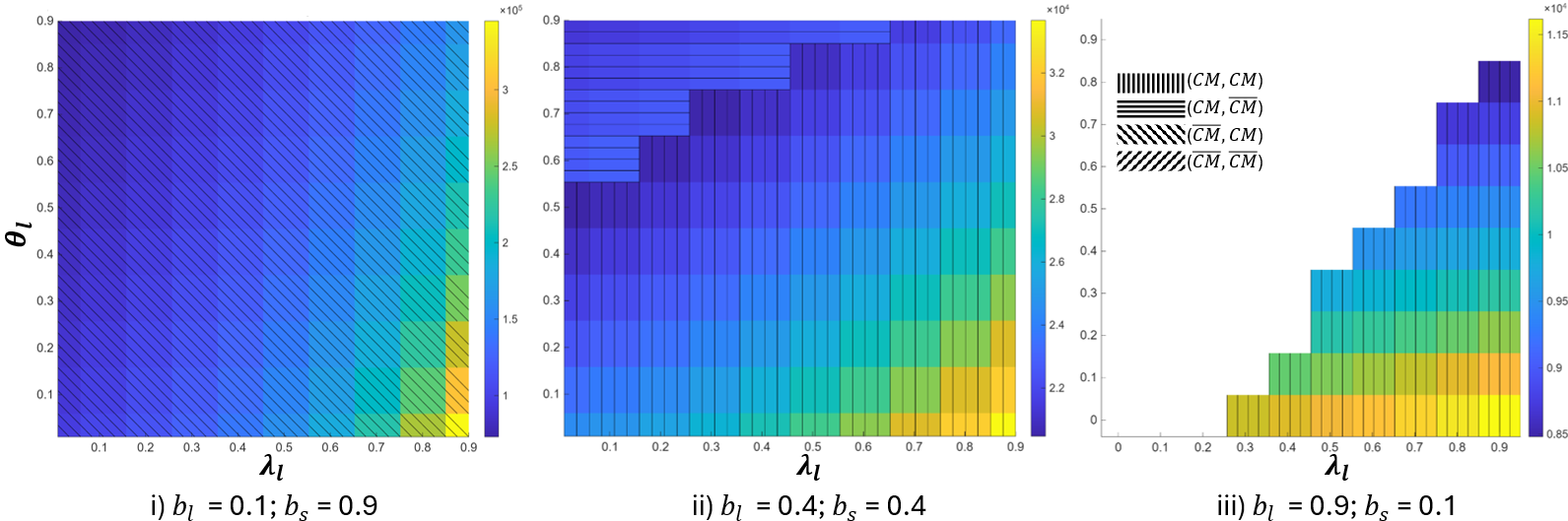}
    \vspace{-1em}
    \caption{$r_1$’s profit from bundling relative to no bundling, measured by $\Delta \pi_{r1}^B$, as a function of $\lambda_l$ and $\theta_l$, for varying $b_l$ and $b_s$. Shaded regions indicate the optimal PMG regime under bundling.}
    \label{fig:lmb_tht_consolidated}
    \vspace{-1em}
\end{figure*}

\begin{figure*}[t]
    \centering
    \includegraphics[width=0.81\linewidth]{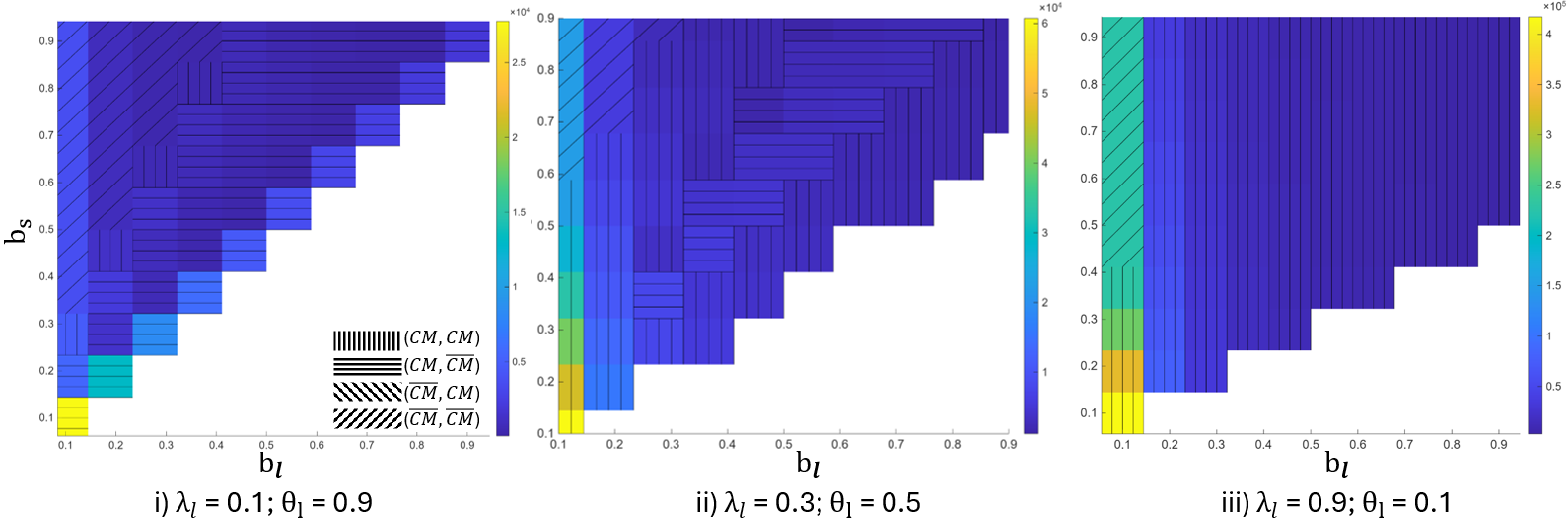}
    \vspace{-1em}
    \caption{$r_1$’s profit from bundling relative to no bundling, measured by $\Delta \pi_{r1}^B$, as a function of $b_l$ and $b_s$, for varying $\lambda_l$ and $\theta_l$. Shaded regions indicate the optimal PMG regime under bundling.}
    \label{fig:bl_bs_consolidated}
    \vspace{-1.5em}
\end{figure*}

\vspace{-0.5em}
\noindent 
While the equilibrium analysis yields closed-form solutions, the interaction of multiple customer segments, mixed bundling, and PMGs makes policy comparisons and selections difficult to characterize analytically. We therefore conduct numerical experiments to (i) identify regions in which bundling and PMG adoption are optimal, and (ii) illustrate how key demand parameters shape equilibrium pricing and profits. Accordingly, our numerical experiments are designed to address the following three related questions: 
\vspace{-0.2em}
\begin{itemize}
    \item {\bf Q1:} When is mixed bundling more profitable for $r_1$ than no bundling?
    \vspace{-0.6em}
    \item {\bf Q2:} When is PMG adoption optimal for $r_1$ conditional on bundling?
    \vspace{-0.6em}
    \item {\bf Q3:} How does the joint choice of bundling and PMGs vary across demand environments?
\end{itemize}
\vspace{-0.2em}
The baseline parameter values follow the bundling literature (see Giri et al., 2020; Taleizadeh et al., 2020), with normalizations where necessary to match our demand system and notation. Unless otherwise stated, we set $a_l^{i1} = a_l^{i2} = a_l^{ib} = a_q^{ib} = a_l^{jb} = a_q^{jb} = a_s = 100, b_l = b_s = 0.4, c_1 = c_2 = 10,$ and $\alpha = 0.5$. For each parameter configuration, we compute the candidate equilibrium prices using the closed-form expressions derived in Section 4 for all relevant subgames. We then retain only those parameter points that satisfy the model's feasibility requirements, including non-negative demands for all active segments and the concavity conditions. Note that when multiple feasible equilibria arise, we select the equilibrium that maximizes $r_1$'s profit.\\
[0.2cm]
We conduct two sets of experiments and present representative parameter slices. While the precise region boundaries may shift with alternative values, the qualitative patterns remain robust across the admissible parameter space. In the first set, we vary the complementarity parameters $(\lambda_l,\theta_l)$ while holding other parameters at baseline levels. We do so under several representative configurations of $(b_s,b_l)$, which capture different price-sensitivities of strategic and loyal customers.
For each parameter combination, we compute $r_1$’s equilibrium profits under no bundling and bundling, where the latter is the maximum equilibrium profit across all bundling subgames. We record the profit difference $\Delta\pi_{r1}^B=\pi_{r1}(B=1)-\pi_{r1}(B=0)$ and identify the corresponding optimal PMG regime, thereby jointly characterizing bundling profitability and PMG adoption.\\
[0.2cm]
In the second set of experiments, we vary $(b_s,b_l)$ under several representative $(\lambda_l,\theta_l)$ configurations. As before, for each $(b_s,b_l)$ pair, we compute $r_1$’s equilibrium profits under no bundling and bundling, identify the profit-maximizing policy, and record the associated PMG regime. This experiment highlights how heterogeneity in customer price-sensitivities shapes both bundling profitability and joint bundling-PMG strategies.\\
[0.2cm]
Figures~\ref{fig:lmb_tht_consolidated} and \ref{fig:bl_bs_consolidated} visualize the numerical experiments' outcomes, reporting $r_1$’s equilibrium profit from bundling relative to no bundling across representative parameter configurations, together with the corresponding optimal PMG regimes. We next summarize the main patterns from these figures, beginning with the profitability of bundling relative to no bundling and then turning to PMG adoption and joint policy choices. The legend and shading scheme are consistent across panels, with blank regions indicating parameter values for which no equilibrium exists.
\vspace{-0.3em}
\begin{observation}\label{Q1_1}
    Mixed bundling ($B = 1$) yields strictly higher profit for $r_1$ than the no-bundling benchmark ($B = 0$) whenever an equilibrium exists. This dominance holds across variations in bundle-level complementarity $\lambda_l$, item-level complementarity $\theta_l$, and own-price sensitivities $(b_l,b_s)$.
\end{observation}
\vspace{-0.3em}
\noindent
Mixed bundling allows $r_1$ to set a separate bundle price for customers who purchase both products, whereas without bundling such demand is priced only through the sum of individual-item prices, limiting $r_1$’s pricing flexibility. As shown in the equilibrium analysis, the availability of a bundle price gives $r_1$ direct control over joint-demand customers, with bundle demand responding to this price through $\lambda_l$. In contrast, under no bundling, joint demand is influenced only indirectly via individual-item prices. Across all feasible parameter configurations, this additional degree of pricing control strictly increases $r_1$’s equilibrium profit whenever an equilibrium exists.
\vspace{-0.3em}
\begin{observation}\label{Q1_2}
    The profit advantage of bundling is largest when bundle-level complementarity $\lambda_l$ is sufficiently high and item-level complementarity $\theta_l$ is sufficiently low. As $\theta_l$ increases, the incremental profit from bundling relative to no bundling decreases.
\end{observation}
\vspace{-0.3em}
\noindent
When bundle-level complementarity $\lambda_l$ is high, bundle demand responds strongly to the bundle price, which increases the value of pricing joint demand separately. By contrast, a higher item-level complementarity $\theta_l$ strengthens the cross-effects between individual-item prices, making joint demand more responsive to individual-item pricing alone. Consequently, the incremental profit gain from bundling relative to no bundling decreases as $\theta_l$ increases, although bundling remains profitable. We next turn to PMG profitability conditional on bundling.
\vspace{-0.3em}
\begin{observation}\label{Q2}
    Conditional on bundling ($B = 1$), $r_1$ optimally refrains from offering PMGs when strategic customers are highly price-sensitive (high $b_s$) and loyal customers are relatively price-insensitive (low $b_l$). In contrast, PMGs become optimal for $r_1$ at medium-to-high values of $(b_l,b_s)$.
\end{observation}
\vspace{-0.3em}
\noindent
When bundling is offered, PMGs link $r_1$’s effective bundle price for price-aware customers to the rival’s bundle price. While this enables $r_1$ to attract strategic customers, the matched lower price is also extended to loyal, price-aware customers, who would purchase from $r_1$ regardless of its price-matching policy. When loyal customers are relatively price-insensitive (low $b_l$), this creates a substantial margin loss, making PMGs unattractive. This happens especially when strategic demand is highly price-sensitive (high $b_s$), which intensifies price competition. As $b_l$ increases, loyal customers become more price-sensitive (i.e., have lower willingness-to-pay), which reduces the margin loss and allows the benefit of strategic demand-capture to dominate, making PMGs optimal.\\
[0.2cm]
Together, Observations \ref{Q1_1}-\ref{Q2} characterize the joint choice of bundling and PMGs across demand environments. Within the admissible parameter space, no bundling does not emerge as optimal; mixed bundling is preferred whenever an equilibrium exists. Conditional on bundling, PMG adoption varies systematically with customer price-sensitivities and complementarity, switching on or off depending on the relative strength of strategic versus loyal demand. Thus, bundling is the primary strategic lever for $r_1$, while PMGs act as a demand-dependent adjustment governing how price-aware customers are served.
\vspace{-0.3em}
\begin{observation}\label{EqExistence}
    Equilibrium fails to exist in regions characterized by strong asymmetries in price-sensitivity, specifically, when loyal customers are highly price-sensitive ($b_l$ very high), and strategic customers are weakly price-sensitive ($b_s$ very low) (blank areas in Figures~\ref{fig:lmb_tht_consolidated} and \ref{fig:bl_bs_consolidated}). Additionally, equilibrium existence is more robust when bundle-level complementarity $\lambda_l$ is sufficiently high and item-level complementarity $\theta_l$ is sufficiently low.
\end{observation}
\vspace{-0.3em}
\noindent
When loyal customers are highly price-sensitive while strategic customers are relatively price-insensitive, pricing responses become highly asymmetric across segments. In this case, small price changes induce large demand adjustments from loyal customers but only weak responses from strategic customers. Such asymmetries can violate the concavity and the conditions required for equilibrium existence. Equilibrium existence is more robust when bundle-level complementarity $\lambda_l$ is sufficiently high and item-level complementarity $\theta_l$ is sufficiently low, as a stronger bundle responsiveness stabilizes joint-demand pricing and preserves the required regularity conditions.
\vspace{-2.1em}
\section{Conclusions}\label{sec6}
\vspace{-0.5em}
\noindent
This paper links two widely used retail levers, including bundling and PMGs, in a competitive market for complementary goods. We develop a duopoly model where Retailer 1 chooses between mixed bundling and no bundling, while Retailer 2 sells a bundle only. Both retailers set prices and may offer PMGs. Our main insight is that under mixed bundling, a PMG can transmit competitive conditions across retailers. In particular, when Retailer 1 offers a PMG, its equilibrium bundle price (and, through bundle-item interactions, its component prices) depends on the rival’s bundle-demand conditions because the effective bundle price faced by price-aware customers is tied to the competitor’s price. However, Retailer 2’s equilibrium bundle price depends only on its own demand. Numerically, mixed bundling strictly dominates no bundling whenever an equilibrium exists. Conditional on bundling, PMG adoption follows a clear trade-off between capturing strategic demand and conceding margins on loyal, price-aware customers, varying systematically with customer price-sensitivities and underlying demand conditions.

\end{document}


\begin{center}
{\large\bfseries Bundling and Price-Matching in Competitive Complementary Goods Markets}
\end{center}
\section*{Supplementary Material (Appendices)} \label{appendix}
\noindent 
Throughout the appendices, multiple sufficient condition sets are introduced to select among competing equilibrium price-ordering regimes. These conditions are not necessary and do not exhaust the parameter space, but are designed to establish existence and conditional uniqueness of equilibrium outcomes.

\begin{notation}
For brevity, we define the following notation throughout this supplementary material:
\[
   t_1:= (b_l+\lambda_l), \ t_2:=(b_l\theta_l+\lambda_l)  
\]
\end{notation}

\subsection*{\bf Appendix A: Proof of Theorem 1}
\noindent 
We first prove this result for the case in which both retailers offer PMGs; the other cases can be analyzed similarly. There are two possible subcases within this setting.
\begin{itemize}
    \item Subcase 1: $p_{r1}^b \ge p_{r2}^{b}$. Then, we have $\Tilde{p}_{r_1}^b = p_{r2}^{b}$, $\Tilde{p}_{r_2}^b = p_{r2}^{b}$, and $\hat{p}^b = p_{r2}^{b}$. Therefore, we have: 
    \begin{subequations}
    \begin{align}
        & \pi_{r1} = (p_{r1}^{i1} - c_1)d_l^{i1} + (p_{r1}^{i2} - c_2)d_l^{i2} + (p_{r1}^b - c_1 -c_2)d_l^{ib}   + ({p_{r2}^{b}} \!-\! c_1 \!-\!c_2)d_q^{ib} + \alpha(p_{r2}^{b}\!-\!c_1\!-\!c_2)d_s \label{profit_r1_case1_clean}\\
        & \pi_{r2} = (p_{r2}^{b} - c_1 -c_2)d_l^{jb} + (p_{r2}^{b} - c_1 -c_2)d_q^{jb} + (1-\alpha)(p_{r2}^{b}-c_1-c_2)d_s \label{profit_r2_case1_clean}
    \end{align}
    \end{subequations}
    Note that the profit terms associated with loyal, price-aware customers in the profit function in~\eqref{profit_r1_case1_clean} also depend on $p_{r2}^{b}$. Moreover, non-loyal, price-aware (strategic) customers are indifferent between the two retailers as they both offer price-matching. Writing the Hessian matrix H (the second-order conditions):
   \begin{subequations}
        \begin{align}
            &\nabla_p^2 \pi_{r1} 
            = 2 \begin{bmatrix} -t_1 & -t_2 & \lambda_l \\ 
            -t_2 & -t_1 & \lambda_l \\
            \lambda_l & \lambda_l & -t_1\\
            \end{bmatrix}, \  
             \nabla_p^2 \pi_{r2} = -4b_l-2(1-\alpha)b_s  < 0 \nonumber 
        \end{align}
    \end{subequations}
     Further, the eigenvalues $e_i, i\in\{1,2,3\}$, of the symmetric Hessian matrix in the above equation are: 
    \begin{subequations}
    \begin{align}
        & e_1 \!=\! -2b_l(1\!-\!\theta_l), \ e_2 \!=\!  - 2b_l \!-\! 3\lambda_l \!-\! b_l\theta_l \!-\! \sqrt{b_l^2 \theta_l^2 \!+\! 2b_l \lambda_l \theta_l \!+\! 9 \lambda_l^2}, \ e_3 
        \!=\!  - 2b_l \!-\! 3\lambda_l \!-\! b_l\theta_l \!+\! \sqrt{(b_l \theta_l \!+\! \lambda_l)^2 \!+\! 8 \lambda_l^2}  \nonumber
    \end{align}
    \end{subequations}
    For $b_l\ge \lambda_l$ and $\theta_l\in [0,1]$, the eigenvalues are all negative. Hence, the symmetric Hessian matrix is negative definite, implying that the profit functions are concave. Writing the first-order conditions, we have:
    \begin{small}
    \begin{subequations}\label{asymm_eqbm_1_first_order_clean}
    \begin{align}
        & \frac{\partial \pi_{r1}}{\partial p_{r1}^{i1}} = \left(d_l^{i1} + (p_{r1}^{i1} - c_1)(-b_l-\lambda_l) + (p_{r1}^{i2} - c_2)(-b_l\theta_l-\lambda_l)\right) + (p_{r1}^b - c_1 - c_2)\lambda_l + (p_{r2}^{b} - c_1 - c_2)\lambda_l = 0  \\ 
        & \frac{\partial \pi_{r1}}{\partial p_{r1}^{i2}} = \left((p_{r1}^{i1} - c_1)(-b_l\theta_l-\lambda_l)+ d_l^{i2} + (p_{r1}^{i2} - c_2)(-b_l-\lambda_l)\right) + (p_{r1}^b - c_1 - c_2)\lambda_l + (p_{r2}^{b} - c_1 - c_2)\lambda_l   = 0 \\
        & \frac{\partial \pi_{r1}}{\partial p_{r1}^b} = (p_{r1}^{i1} - c_1)\lambda_l+ (p_{r1}^{i2} - c_2)\lambda_l + d_l^{ib}  +  (p_{r1}^b - c_1 - c_2)(-b_l-\lambda_l) = 0 \\
        & \frac{\partial \pi_{r2}}{\partial p_{r2}^{b}} = d_l^{jb}+d_q^{jb} + (1-\alpha)d_s + (p_{r2}^{b} - c_1 - c_2)(-2b_l-(1-\alpha)b_s) = 0
    \end{align}
    \end{subequations}
    \end{small} 
    
    \item Subcase 2:  $p_{r1}^b \le p_{r2}^{b}$. Then, we have $\Tilde{p}_{r_1}^b = \Tilde{p}_{r_2}^b = p_{r1}^b$ and $\hat{p}^b = p_{r1}^b$. Therefore, we have:
    \begin{subequations}
    \begin{align}
        & \pi_{r1} = (p_{r1}^{i1} - c_1)d_l^{i1} + (p_{r1}^{i2} - c_2)d_l^{i2} + (p_{r1}^b - c_1 -c_2)d_l^{ib}  + ({p_{r1}^b} - c_1 -c_2)d_q^{ib} + \alpha(p_{r1}^b-c_1-c_2)d_s  \nonumber \\
        & \pi_{r2} = (p_{r2}^{b} - c_1 -c_2)d_l^{jb} + (p_{r1}^b - c_1 -c_2)d_q^{jb} + (1-\alpha)(p_{r1}^b-c_1-c_2)d_s \nonumber 
    \end{align}
    \end{subequations}
        Similarly, writing the Hessian matrix H (the second-order conditions) and calculating the eigenvalues $e_i, i\in\{1,2,3\}$, of the symmetric Hessian matrix, we show that the profit functions are concave. We then write the first-order conditions for equilibrium. 
 \end{itemize}
Solving the first-order conditions yields a unique stationary point. This stationary point is a valid equilibrium provided that the implied prices satisfy $p_{r1}^b \ge p_{r2}^b$ in Subcase 1 and that all demand quantities are nonnegative. Similarly, feasibility of the solution in Subcase 2 requires that the implied prices satisfy $p_{r1}^b \le p_{r2}^b$ and that all demand quantities be nonnegative.\\ 
[0.2cm]
The above feasibility requirements are endogenous and parameter-dependent. In general, the feasibility regions associated with Subcases 1 and 2 may overlap, and neither subcase can be ruled out a priori. We therefore characterize a set of sufficient conditions under which the feasibility conditions of Subcase 1 are satisfied, while those of Subcase 2 are violated.\\
[0.2cm]
\textit{Sufficient Condition Set A.} The following inequalities constitute a set of sufficient conditions:
\begin{align*}
a_l^{i1} + a_l^{i2} &\ge \frac{4}{3}\left(a_l^{jb}+a_q^{jb}\right),\ \  
a_l^{ib} \ge \frac{a_l^{jb}+a_q^{jb}}{2},\ \ 
a_q^{ib} \ge a_q^{jb} \ge a_l^{jb},\\
\frac{a_l^{i1}+a_l^{i2}}{2a_s} &\ge \frac{b_l}{b_s} \ge \frac{a_l^{jb}+a_q^{jb}}{2a_s},\ \ 
\frac{a_l^{ib}}{a_s} \ge \frac{b_l}{b_s} \ge \frac{a_l^{jb}}{a_s}.
\end{align*}
These conditions are sufficient to ensure that the stationary solution derived under Subcase 1 satisfies its defining price-ordering and nonnegativity constraints. At the same time, under Sufficient Condition Set A, the stationary solution associated with Subcase 2 fails to satisfy at least one of its defining feasibility requirements and is therefore not admissible as an equilibrium. 
Hence, a unique equilibrium exists under Sufficient Condition Set A. The proof for $(CM,\overline{CM})$ follows the same steps.

\subsection*{\bf Appendix B: Proof of Theorem 2}
\noindent
We first establish the result for the case in which retailer 1 does not offer a PMG, while Retailer 2 does, i.e., $(\overline{CM}, CM)$. The proofs for the remaining cases follow analogously. Within this case, two subcases arise:
\begin{itemize}
    \item Subcase 1: $p_{r1}^b \ge p_{r2}^{b}$. Then, we have $\Tilde{p}_{r1}^b = p_{r1}^{b}$, $\Tilde{p}_{r2}^b = p_{r2}^{b}$, and $\hat{p}^b = p_{r2}^{b}$. Therefore, we have:
    \begin{subequations}
    \begin{align}
        & \pi_{r1} = (p_{r1}^{i1} - c_1)d_l^{i1} + (p_{r1}^{i2} - c_2)d_l^{i2} + (p_{r1}^b - c_1 -c_2)d_l^{ib}   + ({p_{r1}^{b}} \!-\! c_1 \!-\!c_2)d_q^{ib}  \nonumber
        \\
        & \pi_{r2} = (p_{r2}^{b} - c_1 -c_2)d_l^{jb} + (p_{r2}^{b} - c_1 -c_2)d_q^{jb} + (p_{r2}^{b}-c_1-c_2)d_s \nonumber 
    \end{align}
    \end{subequations}
    Writing the Hessian matrix H (the second-order conditions):
   \begin{subequations}
        \begin{align}
            &\nabla_p^2 \pi_{r1} 
            = 2 \begin{bmatrix} -t_1 & -t_2 & \frac{3}{2}\lambda_l \\ 
            -t_2 & -t_1 & \frac{3}{2}\lambda_l \\
            \frac{3}{2}\lambda_l & \frac{3}{2}\lambda_l & -2t_1\\
            \end{bmatrix}, \ 
            \nabla_p^2 \pi_{r2} = -4b_l-2b_s  < 0  \nonumber
        \end{align}
    \end{subequations}
     Further, calculating the eigenvalues $e_i, i\in\{1,2,3\}$, of the symmetric Hessian matrix,  
    we have: 
    \begin{subequations}
    \begin{align}
        & e_1 \!=\! -2b_l(1\!-\!\theta_l),  
        e_2 \!=\!  - 3b_l \!-\! 4\lambda_l \!-\! b_l\theta_l -\!\! \sqrt{b_l^2 (1\!-\!\theta_l)^2 \!+\! 18 \lambda_l^2}, \ e_3 
        =  - 3b_l \!-\! 4\lambda_l \!-\! b_l\theta_l \!+\!\! \sqrt{b_l^2 (1\!-\!\theta_l)^2 \!+\! 18 \lambda_l^2} \nonumber 
    \end{align}
    \end{subequations}
    For $b_l\ge \lambda_l$ and $\theta_l\in [0,1]$, the eigenvalues are all negative. 
    Hence, the profit functions are concave. Writing the first-order conditions for equilibrium, we have:
    \begin{small}
    \begin{subequations}\label{asymm_eqbm_1_first_order}
    \begin{align}
        & \frac{\partial \pi_{r1}}{\partial p_{r1}^{i1}} = \left(d_l^{i1} + (p_{r1}^{i1} - c_1)(-b_l-\lambda_l) + (p_{r1}^{i2} - c_2)(-b_l\theta_l-\lambda_l)\right)  + (p_{r1}^b - c_1 - c_2)\lambda_l + (p_{r1}^b - c_1 - c_2)\lambda_l = 0  \\ 
        & \frac{\partial \pi_{r1}}{\partial p_{r1}^{i2}} = \left((p_{r1}^{i1} - c_1)(-b_l\theta_l-\lambda_l)+ d_l^{i2} + (p_{r1}^{i2} - c_2)(-b_l-\lambda_l)\right)  + (p_{r1}^b - c_1 - c_2)\lambda_l + (p_{r1}^b - c_1 - c_2)\lambda_l   = 0 \\
        & \frac{\partial \pi_{r1}}{\partial p_{r1}^b} = (p_{r1}^{i1} - c_1)\lambda_l+ (p_{r1}^{i2} - c_2)\lambda_l + d_l^{ib} + d_q^{ib}  +  (p_{r1}^b - c_1 - c_2)(-2b_l-2\lambda_l) = 0 \\
        & \frac{\partial \pi_{r2}}{\partial p_{r2}^{b}} = d_l^{jb}+d_q^{jb} + d_s + (p_{r2}^{b} - c_1 - c_2)(-2b_l-b_s) = 0
    \end{align}
    \end{subequations}
    \end{small} 
    
    \item Subcase 2:  $p_{r1}^b \le p_{r2}^{b}$. Then, we have $\Tilde{p}_{r_1}^b = \Tilde{p}_{r_2}^b = p_{r1}^b$ and $\hat{p}^b = p_{r1}^b$. Therefore, we have:
    \begin{subequations}
    \begin{align}
        & \pi_{r1} = (p_{r1}^{i1} - c_1)d_l^{i1} + (p_{r1}^{i2} - c_2)d_l^{i2} + (p_{r1}^b - c_1 -c_2)d_l^{ib}  + ({p_{r1}^b} - c_1 -c_2)d_q^{ib} + \alpha(p_{r1}^b-c_1-c_2)d_s  \label{profit_r1_case1_sub2}\\
        & \pi_{r2} = (p_{r2}^{b} - c_1 -c_2)d_l^{jb} + (p_{r1}^b - c_1 -c_2)d_q^{jb} + (1-\alpha)(p_{r1}^b-c_1-c_2)d_s\label{profit_r2_case1_sub2}
    \end{align}
    \end{subequations}
        Similarly, writing the Hessian matrix H (the second-order conditions) and calculating the eigenvalues $e_i, i\in\{1,2,3\}$, of the symmetric Hessian matrix, we show that the profit functions are concave. We then solve the first-order conditions for equilibrium.
    \end{itemize}
Solving the first-order conditions yields a unique stationary point. This stationary point is a valid equilibrium provided that the implied prices satisfy $p_{r1}^b \ge p_{r2}^b$ in Subcase 1 and that all demand quantities are nonnegative. Similarly, feasibility of the solution in Subcase 2 requires that the implied prices satisfy $p_{r1}^b \le p_{r2}^b$ and that all demand quantities be nonnegative. \\
[0.2cm]
The above feasibility requirements are endogenous and parameter-dependent. In general, the feasibility regions associated with Subcases 1 and 2 may overlap, and neither subcase can be ruled out a priori. We therefore characterize a set of sufficient conditions under which the feasibility conditions of Subcase 1 are satisfied, while those of Subcase 2 are violated.\\
[0.2cm] 
\textit{Sufficient Condition Set B.} The following inequalities constitute a set of sufficient conditions:
    \begin{align*}
        a_l^{i1} + a_l^{i2} &\ge \frac{4}{3}\left(a_l^{jb}+a_q^{jb}\right),\ \ 
        a_l^{ib} + a_q^{ib} \ge a_l^{jb}+a_q^{jb},\ \ 
        a_q^{jb} \ge a_l^{jb},\\
        \frac{a_l^{i1}+a_l^{i2}}{2a_s} &\ge \frac{4}{3}\frac{b_l}{b_s} \ge \frac{a_q^{jb}+a_q^{jb}}{2a_s},\ \
        \frac{b_l}{b_s} \ge \frac{a_l^{jb}}{a_s}.
    \end{align*}
These conditions are sufficient to ensure that the stationary solution derived under Subcase 1 satisfies its defining price-ordering and nonnegativity constraints. At the same time, under Sufficient Condition Set B, the stationary solution associated with Subcase 2 fails to satisfy at least one of its defining feasibility requirements and is therefore not admissible as an equilibrium. Hence, a unique equilibrium exists under Sufficient Condition Set B. The proof for the strategy $(\overline{CM}, \overline{CM})$ follows the same structure.



\subsection*{\bf Appendix C: Proof of Theorem 3}
\noindent
The price faced by loyal, price-aware customers varies with the retailers’ price-matching preferences. We first establish the result for the case in which both retailers offer price-matching; the other cases follow analogously. There are two possible Subcases in this setting:
\begin{itemize}
    \item Subcase 1: $p_{r1}^b \le p_{r2}^{b}$. Then, we have $\Tilde{p}_{r_1}^b = p_{r1}^{b}$, $\Tilde{p}_{r_2}^b = p_{r1}^{b}$, and $\hat{p}^b = p_{r1}^{b}$. Therefore, we have:
    \begin{subequations}
    \begin{align}
        & \pi_{r1} = (p_{r1}^{i1} - c_1)d_l^{i1} + (p_{r1}^{i2} - c_2)d_l^{i2} + (p_{r1}^b - c_1 -c_2)d_l^{ib}   + ({p_{r1}^{b}} \!-\! c_1 \!-\!c_2)d_q^{ib} + \alpha(p_{r1}^{b}\!-\!c_1\!-\!c_2)d_s \label{profit_r1_case3_clean}\\
        & \pi_{r2} = (p_{r2}^{b} - c_1 -c_2)d_l^{jb} + (p_{r1}^{b} - c_1 -c_2)d_q^{jb} + (1-\alpha)(p_{r1}^{b}-c_1-c_2)d_s \label{profit_r2_case3_clean}
    \end{align}
    \end{subequations}
   Writing the Hessian matrix H (the second-order conditions):
   \begin{subequations}
        \begin{align}
            &\nabla_p^2 \pi_{r1} 
            = 2 \begin{bmatrix} -t_1 & -t_2 & \frac{3}{2}\lambda_l \\ 
            -t_2 & -t_1 & \frac{3}{2}\lambda_l \\
            \frac{3}{2}\lambda_l & \frac{3}{2}\lambda_l & -2t_1-\alpha b_s\\
            \end{bmatrix}, \ 
            \nabla_p^2 \pi_{r2} = -2b_l < 0 
        \end{align}
    \end{subequations}
     Further, calculating the eigenvalues $e_i, i\in\{1,2,3\}$, of the symmetric Hessian matrix above yields: 
    \begin{subequations}
    \begin{align}
        & e_1 \!=\! -2b_l(1\!-\!\theta_l), \ e_2 \!=\!  -\alpha b_s - b_l\theta_l - 3b_l - 4\lambda_l - \psi, \ e_3 
        \!=\!  -\alpha b_s - b_l\theta_l - 3b_l - 4\lambda_l + \psi  \nonumber
    \end{align}
    \end{subequations}
    where $\psi:= \sqrt{(\alpha b_s)^2 - 2\alpha b_sb_l\theta_l + 2\alpha b_sb_l + b_l^2\theta_l^2 - 2b_l^2\theta_l + b_l^2 + 18\lambda_l^2}$. For $b_l\ge \lambda_l$ and $\theta_l\in [0,1]$, all eigenvalues are negative. Writing the first-order conditions for equilibrium, we have:
    \begin{small}
    \begin{subequations}\label{asymm_eqbm_3_first_order_clean}
    \begin{align}
        & \frac{\partial \pi_{r1}}{\partial p_{r1}^{i1}} = \left(d_l^{i1} + (p_{r1}^{i1} - c_1)(-b_l-\lambda_l) + (p_{r1}^{i2} - c_2)(-b_l\theta_l-\lambda_l)\right) + (p_{r1}^b - c_1 - c_2)\lambda_l + (p_{r1}^{b} - c_1 - c_2)\lambda_l = 0  \\ 
        & \frac{\partial \pi_{r1}}{\partial p_{r1}^{i2}} = \left((p_{r1}^{i1} - c_1)(-b_l\theta_l-\lambda_l)+ d_l^{i2} + (p_{r1}^{i2} - c_2)(-b_l-\lambda_l)\right) + (p_{r1}^b - c_1 - c_2)\lambda_l + (p_{r1}^{b} - c_1 - c_2)\lambda_l   = 0 \\
        & \frac{\partial \pi_{r1}}{\partial p_{r1}^b} = (p_{r1}^{i1} - c_1)\lambda_l+ (p_{r1}^{i2} - c_2)\lambda_l + d_q^{ib} + d_l^{ib}  +  (p_{r1}^b - c_1 - c_2)(-2b_l-2\lambda_l-\alpha b_s) + \alpha d_s= 0 \\
        & \frac{\partial \pi_{r2}}{\partial p_{r2}^{b}} = d_l^{jb} - (p_{r2}^{b} - c_1 - c_2)b_l = 0
    \end{align}
    \end{subequations}
    \end{small} 
    
    \item Subcase 2:  $p_{r1}^b \ge p_{r2}^{b}$. Then, we have $\Tilde{p}_{r_1}^b = \Tilde{p}_{r_2}^b = p_{r2}^b$ and $\hat{p}^b = p_{r2}^b$. Therefore, we have:
    \begin{subequations}
    \begin{align}
        & \pi_{r1} = (p_{r1}^{i1} - c_1)d_l^{i1} + (p_{r1}^{i2} - c_2)d_l^{i2} + (p_{r1}^b - c_1 -c_2)d_l^{ib}  + ({p_{r2}^b} - c_1 -c_2)d_q^{ib} + \alpha(p_{r2}^b-c_1-c_2)d_s  \nonumber \\
        & \pi_{r2} = (p_{r2}^{b} - c_1 -c_2)d_l^{jb} + (p_{r2}^b - c_1 -c_2)d_q^{jb} + (1-\alpha)(p_{r2}^b-c_1-c_2)d_s \nonumber 
    \end{align}
    \end{subequations}
Similarly, writing the Hessian matrix H (the second-order conditions) and calculating the eigenvalues $e_i, i\in\{1,2,3\}$, of the symmetric Hessian matrix, we show that the profit functions are concave. We then solve the first-order conditions for equilibrium. 
\end{itemize}
Solving the first-order conditions yields a unique stationary point. This stationary point is a valid equilibrium provided that the implied prices satisfy $p_{r1}^b \le p_{r2}^b$ in Subcase 1 and that all demand quantities are nonnegative. Similarly, feasibility of the solution in Subcase 2 requires that the implied prices satisfy $p_{r1}^b \ge p_{r2}^b$ and that all demand quantities be nonnegative.\\ 
[0.2cm]
The above feasibility requirements are endogenous and parameter-dependent. In general, the feasibility regions associated with Subcases 1 and 2 may overlap, and neither subcase can be ruled out a priori. We therefore characterize a set of sufficient conditions under which the feasibility conditions of Subcase 1 are satisfied, while those of Subcase 2 are violated.\\
[0.2cm]
\textit{Sufficient Condition Set C.} The following inequalities constitute a set of sufficient conditions:
        \begin{align*}
            \frac{a_l^{i1}+a_l^{i2}}{(1+\theta_l) a_s} &\le \frac{b_l}{b_s} \le \frac{a_l^{jb}+a_q^{jb}}{2a_s},\ \ \frac{a_l^{ib}}{a_s} \frac{2}{1+\theta_l} \le \frac{b_l}{b_s} \le \frac{a_l^{jb}}{a_s}, \\
            a_l^{ib} \frac{2}{1+\theta_l} &\le \frac{a_l^{jb}+a_q^{jb}}{2},\ \
            a_q^{ib} \le a_l^{ib}  \le a_q^{jb} \le a_l^{jb},\\
            (a_l^{ib}+a_q^{ib}) & + \frac{c_1+c_2}{2}\frac{2b_l+b_s}{2b_l}\lambda_l  \le a_l^{jb}+a_q^{jb}, \\
            a_l^{jb} & \le (c_1+c_2)b_l, \ \
            a_s  \le (c_1+c_2)(1-\alpha)b_s
        \end{align*}
These conditions are sufficient to ensure that the stationary solution derived under Subcase 1 satisfies its defining price-ordering and nonnegativity constraints. At the same time, under Sufficient Condition Set C, the stationary solution associated with Subcase 2 fails to satisfy at least one of its defining feasibility requirements and is therefore not admissible as an equilibrium. 
Hence, a unique equilibrium exists under Sufficient Condition Set C. The proof follows the same structure for the strategy $( \overline{CM},CM)$.



\subsection*{\bf Appendix D: Proof of Theorem 4}
\noindent
We first establish the result for the case in which Retailer 1 does offer a PMG, while Retailer 2 does not, i.e., $(CM, \overline{CM})$. The proof for the other cases follows analogously. Within this case, two subcases arise:
\begin{itemize}
    \item Subcase 1: $p_{r1}^b \le p_{r2}^{b}$. Then, we have $\Tilde{p}_{r1}^b = p_{r1}^{b}$ and $\hat{p}^b = p_{r1}^{b}$. Therefore, we have:
    \begin{subequations}
    \begin{align}
        & \pi_{r1} = (p_{r1}^{i1} - c_1)d_l^{i1} + (p_{r1}^{i2} - c_2)d_l^{i2} + (p_{r1}^b - c_1 -c_2)d_l^{ib}   + ({p_{r1}^{b}} \!-\! c_1 \!-\!c_2)d_q^{ib} + (p_{r1}^{b}-c_1-c_2)d_s \nonumber
        \\
        & \pi_{r2} = (p_{r2}^{b} - c_1 -c_2)d_l^{jb} + (p_{r2}^{b} - c_1 -c_2)d_q^{jb}  \nonumber 
    \end{align}
    \end{subequations}
    Writing the Hessian matrix H (the second-order conditions):
   \begin{subequations}
        \begin{align}
            &\nabla_p^2 \pi_{r1} 
            = 2 \begin{bmatrix} -t_1 & -t_2 & \frac{3}{2}\lambda_l \\ 
            -t_2 & -t_1 & \frac{3}{2}\lambda_l \\
            \frac{3}{2}\lambda_l & \frac{3}{2}\lambda_l & -2t_1-b_s\\
            \end{bmatrix}, \ 
            \nabla_p^2 \pi_{r2} = -4b_l  < 0  \nonumber
        \end{align}
    \end{subequations}
     Further, calculating the eigenvalues $e_i, i\in\{1,2,3\}$, of the symmetric Hessian matrix yields:  
    \begin{subequations}
    \begin{align}
        & e_1 \!=\! -2b_l(1\!-\!\theta_l),  
        e_2 \!=\!   -b_s - b_l\theta_l - 3b_l - 4\lambda_l -\psi, \ e_3 
        =   -b_s - b_l\theta_l - 3b_l - 4\lambda_l + \psi \nonumber 
    \end{align}
    \end{subequations}
    where $\psi:= \sqrt{b_s^2 - 2b_sb_l\theta_l + 2b_sb_l + b_l^2\theta_l^2 - 2b_l^2\theta_l + b_l^2 + 18\lambda_l^2}$.
    For $b_l\ge \lambda_l$ and $\theta_l\in [0,1]$, all eigenvalues are negative. Hence, the profit functions are concave. Writing the first-order conditions for equilibrium, we have: 
    \begin{small}
    \begin{subequations}\label{asymm_eqbm_4_first_order_clean}
    \begin{align}
        & \frac{\partial \pi_{r1}}{\partial p_{r1}^{i1}} = \left(d_l^{i1} + (p_{r1}^{i1} - c_1)(-b_l-\lambda_l) + (p_{r1}^{i2} - c_2)(-b_l\theta_l-\lambda_l)\right)  + (p_{r1}^b - c_1 - c_2)\lambda_l + (p_{r1}^b - c_1 - c_2)\lambda_l = 0  \\ 
        & \frac{\partial \pi_{r1}}{\partial p_{r1}^{i2}} = \left((p_{r1}^{i1} - c_1)(-b_l\theta_l-\lambda_l)+ d_l^{i2} + (p_{r1}^{i2} - c_2)(-b_l-\lambda_l)\right)  + (p_{r1}^b - c_1 - c_2)\lambda_l + (p_{r1}^b - c_1 - c_2)\lambda_l   = 0 \\
        & \frac{\partial \pi_{r1}}{\partial p_{r1}^b} = (p_{r1}^{i1} - c_1)\lambda_l+ (p_{r1}^{i2} - c_2)\lambda_l + d_l^{ib} + d_q^{ib}  +  (p_{r1}^b - c_1 - c_2)(-2b_l-2\lambda_l- b_s) +   d_s = 0 \\
        & \frac{\partial \pi_{r2}}{\partial p_{r2}^{b}} = d_l^{jb}+d_q^{jb} - 2(p^b_{r2}-c_1-c_2)b_l= 0
    \end{align}
    \end{subequations}
    \end{small} 
    
    \item Subcase 2:  $p_{r1}^b \ge p_{r2}^{b}$. Then, we have $\Tilde{p}_{r_1}^b = p_{r2}^b$ and $\hat{p}^b = p_{r2}^b$. Therefore, we have:
    \begin{subequations}
    \begin{align}
        & \pi_{r1} = (p_{r1}^{i1} - c_1)d_l^{i1} + (p_{r1}^{i2} - c_2)d_l^{i2} + (p_{r1}^b - c_1 -c_2)d_l^{ib}  + ({p_{r2}^b} - c_1 -c_2)d_q^{ib} + \alpha(p_{r2}^b-c_1-c_2)d_s  \label{profit_r1_case1_sub2}\\
        & \pi_{r2} = (p_{r2}^{b} - c_1 -c_2)d_l^{jb} + (p_{r2}^b - c_1 -c_2)d_q^{jb} + (1-\alpha)(p_{r2}^b-c_1-c_2)d_s\label{profit_r2_case1_sub2}
    \end{align}
    \end{subequations}
    Similarly, writing the Hessian matrix H or the second-order condition and calculating the eigenvalues $e_i, i\in\{1,2,3\}$, of the symmetric Hessian matrix, we show that the profit functions are concave. For the equilibrium, we solve the first-order conditions.
    \end{itemize}
Solving the first-order conditions yields a unique stationary point. This stationary point is a valid equilibrium provided that the implied prices satisfy $p_{r1}^b \le p_{r2}^b$ in Subcase 1 and that all demand quantities are nonnegative. Similarly, feasibility of the solution in Subcase 2 requires that the implied prices satisfy $p_{r1}^b \ge p_{r2}^b$ and that all demand quantities be nonnegative. \\
[0.2cm]
The above feasibility requirements are endogenous and parameter-dependent. In general, the feasibility regions associated with Subcases 1 and 2 may overlap, and neither subcase can be ruled out a priori. We therefore characterize a set of sufficient conditions under which the feasibility conditions of Subcase 1 are satisfied, while those of Subcase 2 are violated.\\
[0.2cm]
\textit{Sufficient Condition Set D.} The following inequalities constitute a set of sufficient conditions:
\begin{align*}
\frac{a_l^{i1}+a_l^{i2}}{(1+\theta_l) a_s} &\le \frac{b_l}{b_s} \le \frac{a_l^{jb}+a_q^{jb}}{2a_s},\ \ \frac{a_l^{ib}}{a_s} \frac{2}{1+\theta_l}\le \frac{b_l}{b_s} \le \frac{a_l^{jb}}{a_s}, \\
(a_l^{ib} + a_q^{ib}) \frac{2}{1+\theta_l} & + \frac{c_1+c_2}{2}\frac{2b_l+b_s}{2b_l}\lambda_l \le(a_l^{jb}+a_q^{jb}),\\
a_q^{jb} \le & a_l^{jb},\ \
a_l^{jb}  \le (c_1+c_2)b_l, \ \
a_s \le (c_1+c_2)(1-\alpha)b_s
\end{align*}   
These conditions are sufficient to ensure that the stationary solution derived under Subcase 1 satisfies its defining price-ordering and nonnegativity constraints. At the same time, under Sufficient Condition Set C, the stationary solution associated with Subcase 2 fails to satisfy at least one of its defining feasibility requirements and is therefore not admissible as an equilibrium. 
Hence, a unique equilibrium exists under Sufficient Condition Set D. The proof follows the same structure for the strategy $(\overline{CM}, \overline{CM})$.



\subsection*{\bf Appendix E: Proof of Theorem 5}
\noindent
There are two possible cases determined by the relative price difference between Retailer 1’s combined product and Retailer 2’s bundle:
\begin{itemize}
    \item Case 1: $p_{r1}^{i1} + p_{r1}^{i2} \ge p_{r2}^{b}$. Therefore, 
    \begin{small}
    \begin{subequations}
    \begin{align}
        & \pi_{r1} = (p_{r1}^{i1} - c_1)d_l^{i1} + (p_{r1}^{i2} - c_2)d_l^{i2} + (p_{r1}^{i1}+p_{r1}^{i2} - c_1 -c_2)d_l^{ib}  + (p_{r1}^{i1}+p_{r1}^{i2} - c_1 -c_2)d_q^{ib} \nonumber \\
        & \pi_{r2} \!=\! (p_{r2}^{b} \!-\! c_1 \!-\!c_2)d_l^{jb} \!+\! ({p_b}^{r_2} \!-\! c_1 \!-\!c_2)d_q^{jb} \!+\! (p_{r2}^{b}\!-\!c_1\!-\!c_2)d_s  \nonumber  
    \end{align}
    \end{subequations}
    \end{small}
    Writing the first-order conditions, we have:
    \begin{small}
    \begin{subequations}
    \begin{align}
        \!\!\!\nabla_p \pi_{r1} & \!=\!
            \begin{bmatrix}
                d_l^{i1} \!-\! (p_{r1}^{i1} \!-\! c_1)b_l \!-\! (p_{r1}^{i2} \!-\! c_2)b_l\theta_l
                \!+\! d_l^{ib} \!+\! d_q^{ib} \!-\! 2 (p_{r1}^{i1} \!+\! p_{r1}^{i2} \!-\! c_1 \!-\! c_2)b_l \\[4pt]
                -(p_{r1}^{i1} \!-\! c_1)b_l\theta_l \!+\! d_l^{i2} \!-\! (p_{r1}^{i2} \!-\! c_2)b_l
                \!+\! d_l^{ib} \!+\! d_q^{ib} \!-\! 2 (p_{r1}^{i1} \!+\! p_{r1}^{i2} \!-\! c_1 \!-\! c_2)b_l
            \end{bmatrix}, \
        \nabla_p \pi_{r2} \!=\! d_l^{jb} \!+\! d_q^{jb} \!+\! d_s \!-\! (p_{r2}^{b} \!-\! c_1 \!-\! c_2)(2b_l \!+\! b_s) \nonumber 
    \end{align}
    \end{subequations}
    \end{small}
    Also, from the second-order derivatives, we have:
    \begin{small}
    \begin{subequations}
    \begin{align}
            &\nabla_p^2 \pi_{r1} = -2b_l\begin{bmatrix} 3 & 2+\theta_l \\ 
            2+\theta_l & 3
            \end{bmatrix}, \ \nabla_p^2 \pi_{r2} = -4b_l - 2b_s< 0 \nonumber 
    \end{align}
    \end{subequations}
    \end{small}
    The two eigenvalues of the symmetric Hessian matrix ($\nabla_p^2 \pi_{r1}$) are: 
    \[
        e_1= -2b_l(1-\theta_l), \ e_2 = -2b_l(5+\theta_l)
    \]
    Hence, the optimization problem is concave. We then solve the first-order equations above to obtain the equilibrium. 
    \item Case 2: $p_{r1}^{i1} + p_{r1}^{i2} < p_{r2}^{b}$. Therefore, 
    \begin{small}
    \begin{subequations}
    \begin{align}
        & \pi_{r1} = (p_{r1}^{i1} - c_1)d_l^{i1} + (p_{r1}^{i2} - c_2)d_l^{i2} + (p_{r1}^{i1}+p_{r1}^{i2} - c_1 -c_2)d_l^{ib}  + (p_{r1}^{i1}+p_{r1}^{i2} - c_1 -c_2)d_q^{ib} +\! (p_{r1}^{i1}+p_{r1}^{i2}\!-\!c_1\!-\!c_2)d_s \nonumber \\
        & \pi_{r2} \!=\! (p_{r2}^{b} \!-\! c_1 \!-\!c_2)d_l^{jb} \!+\! (p^b_{r2} \!-\! c_1 \!-\!c_2)d_q^{jb} \nonumber 
    \end{align}
    \end{subequations}
    \end{small}
    Writing the first-order conditions, we have:
    \begin{small}
    \begin{subequations}\label{asymm_eqbm_5_first_order2}
    \begin{align}
        \!\!\!\nabla_p \pi_{r1} & \!=\!
            \begin{bmatrix}
                d_l^{i1} \!-\! (p_{r1}^{i1} \!-\! c_1)b_l \!-\! (p_{r1}^{i2} \!-\! c_2)b_l\theta_l
                \!+\! d_l^{ib} \!+\! d_q^{ib}\!+\!d_s \!-\! (p_{r1}^{i1} \!+\! p_{r1}^{i2} \!-\! c_1 \!-\! c_2)(2b_l\!+\!b_s) \\[4pt]
                -(p_{r1}^{i1} \!-\! c_1)b_l\theta_l \!+\! d_l^{i2} \!-\! (p_{r1}^{i2} \!-\! c_2)b_l
                \!+\! d_l^{ib} \!+\! d_q^{ib} \!+\! d_s \!-\! (p_{r1}^{i1} \!+\! p_{r1}^{i2} \!-\! c_1 \!-\! c_2)(2b_l\!+\!b_s)
            \end{bmatrix}, \
        \nabla_p \pi_{r2} \!=\! d_l^{jb} \!+\! d_q^{jb}
            \!-\! 2(p_{r2}^{b} \!-\! c_1 \!-\! c_2)b_l     \nonumber 
    \end{align}
    \end{subequations}
    \end{small}
    Also, from the second-order derivatives, we have:
    \begin{small}
    \begin{subequations}
    \begin{align}
            &\nabla_p^2 \pi_{r1} = \begin{bmatrix} -6b_l-2b_s & -4b_l-2b_s-2b_l\theta_l \\ 
            -4b_l-2b_s-2b_l\theta_l & -6b_l -2b_s
            \end{bmatrix}, \ \nabla_p^2 \pi_{r2} = -4b_l < 0 
    \end{align}
    \end{subequations}
    \end{small}
    The two eigenvalues of the symmetric Hessian matrix ($\nabla_p^2 \pi_{r1}$) are: 
    \[
        e_1= -2b_l(1-\theta_l), \ e_2 = -10b_l - 4b_s -2b_l\theta_l
    \]
    Hence, the optimization problem is concave. Similarly, we solve the first-order equations above to obtain the equilibrium.
\end{itemize}
Solving the first-order conditions in each case yields a unique stationary point. The stationary point associated with Case 1 constitutes a valid equilibrium provided that the implied prices satisfy $p_{r1}^b \le p_{r2}^b$ and that all demand quantities are nonnegative. Similarly, feasibility of the stationary point in Case 2 requires that the implied prices satisfy $p_{r1}^b \ge p_{r2}^b$ and that all demand quantities be nonnegative.\\
[0.2cm]
The above feasibility requirements are endogenous and depend on the underlying demand and cost parameters. In general, the feasibility regions associated with Cases 1 and 2 need not be disjoint, and neither case can be ruled out a priori. We therefore introduce distinct sets of sufficient conditions to characterize parameter regions under which one case yields an admissible equilibrium while the other does not.\\
[0.2cm]
\textit{Sufficient Condition Set E.} The following inequalities constitute a set of sufficient conditions:
\begin{align*}
a_l^{ib} + a_q^{ib} &\ge a_l^{jb} + a_q^{jb},\ \
a_l^{i1} + a_l^{i2} \ge a_l^{jb} + a_q^{jb},\\
\frac{a_l^{i1} + a_l^{i2}}{2a_s} &\ge \frac{b_l}{b_s},\ \
\frac{a_l^{ib} + a_q^{ib}}{2a_s} \ge \frac{b_l}{b_s} \ge \frac{a_l^{jb} + a_q^{jb}}{2a_s}.
\end{align*}
\textit{Sufficient Condition Set F.} The following inequalities constitute a set of sufficient conditions:
\begin{align*}
a_l^{ib} + a_q^{ib} &\le \frac{3+\theta_l}{4}(a_l^{jb} + a_q^{jb}),\ \
a_l^{i1} + a_l^{i2} \le a_l^{jb} + a_q^{jb},\\
\frac{a_l^{i1} + a_l^{i2}}{2a_s} &\le \frac{b_l}{b_s},\ \
\frac{a_l^{ib} + a_q^{ib}}{2a_s} \frac{4}{3+\theta_l}\le \frac{b_l}{b_s} \le \frac{a_l^{jb} + a_q^{jb}}{2a_s}.
\end{align*}
Under Sufficient Condition Set E, the stationary solution derived under Case 1 satisfies all constraints, while the stationary solution associated with Case 2 fails to satisfy at least one of its feasibility requirements and is therefore not admissible as an equilibrium. Conversely, under Sufficient Condition Set F, the stationary solution derived under Case 2 satisfies its defining feasibility constraints, whereas the stationary solution associated with Case 1 is not admissible. Intuitively, Sufficient Condition Sets E and F govern whether the combined (bundled) price offered by Retailer 1 or the bundled price offered by Retailer 2 is more attractive to price-aware consumers, thereby determining the consistent price-ordering regime in equilibrium.\\
[0.2cm]
Hence, within the strategy profile under consideration, Case 1 constitutes the equilibrium outcome under Sufficient Condition Set E, while Case 2 constitutes the equilibrium outcome under Sufficient Condition Set F.